# Insights into electronic and transport properties of phosphorene nanorings in two perpendicular directions: Effects of circular and elliptical external potentials

M.Amir Bazrafshan[1], Farhad Khoeini*,[1], Bartlomiej Szafran[2]

[1]*Department of Physics, University of Zanjan, P.O. Box 45195-313, Zanjan, Iran*

[2]*Faculty of Physics and Applied Computer Science, AGH University of Science and Technology, al. Mickiewicza 30, 30-059 Kraków, Poland*

**Abstract**

In this work, we study the electronic and transport properties of phosphorene nanorings in two perpendicular directions (zigzag and armchair directions) in the presence of zigzag metallic source and drain leads. Our results are based on the non-equilibrium Green's function (NEGF) method and a five-parameter tight-binding (TB) approach. We investigate how system parameters affect the electronic transport. These parameters include the radius of the rings, the width of the leads and the external potential. Our results show that for all configurations studied, a transport energy gap exists whose width can be tuned by the width of the leads and the radius of the nanoring. The transmission function of wider leads shows more sensitivity to the variation of the inner radius due to higher electronic states that can respond to smaller changes in the scattering region. In addition, the transport along the armchair direction is more susceptible to tuning than the transport along the zigzag direction. The effects of external potentials on the conductance are more pronounced than the geometrical parameters. In particular the circular potential of the amplitude of 0.1 eV can widen the transport gap by about ~0.35 eV.

**Introduction**

One of the most recent and intriguing allotropes of phosphorus is phosphorene, which is a two-dimensional (2D) material derived from black phosphorus in 2014 by mechanical exfoliation [1]. Phosphorene has a puckered honeycomb structure [2–5] with $sp^3$ hybridized phosphorus atoms and exhibits many unique properties, such as high carrier mobility, anisotropic behavior, a tunable band gap, and high flexibility [6–8]. Phosphorene can be manufactured by various methods, such as mechanical exfoliation, liquid exfoliation, chemical vapor deposition, and molecular beam epitaxy [5]. Phosphorene has potential applications in various fields, such as energy storage [9], field-effect transistors [6,10,11], optoelectronic [12], and biosensors [13].

The geometry and shape of nanostructures are crucial for their physical properties [14–18]. For example, the zigzag edge geometry of a graphene nanoribbon makes it a magnetic nanostructure [19]. However, by overcoming the technical issues, the fabrication of nanostructures with more precise dimensions becomes feasible. Phosphorene has been widely studied in the literature. Phosphorene is a direct p-type semiconductor with a gap value of ~1.5 eV [1,20]. A density functional study reports that the direct band gap of the bilayer phosphorene can vary from 0.78 to 1.04 eV depending on the stacking order [21]. Moreover, theoretical studies of zigzag phosphorene nanoribbons (ZPNRs) reveal that they are metallic [22], while armchair phosphorene nanoribbons (APNRs) are all semiconductors [6]. Furthermore, nanoring structures are helpful in studying quantum interference-related effects such as Aharonov-Bohm and Fano resonance [16,23–25]. Ref. [26] reports the possible application of a phosphorene nanoring in sensing biomarker vapors of severe kidney disease. Fano resonance [25] has been studied in a system consisting of bilayer zigzag phosphorene nanoribbons connected to a bilayer phosphorene ring.

In this work, we study the electronic transport properties of a circular phosphorene nanoring with an outer radius ($R_o$) of 6 nm and a range of inner radii ($R_{inner}$) from 2 to 5.5 nm. We show below, that the effect of the outer radius of a circular nanodisk on the transport gap disappears for larger radii, which is a consequence of the semiconducting properties of bulk phosphorene. An anisotropy of the electronic structure and transport is embedded in phosphorene crystal lattice. Therefore, for our study we consider two distinct configurations by connecting two leads in the zigzag and armchair directions. In a transport system, the leads provide/gather electrons to/from the device. Furthermore, we study the effect of lead width, which is determined by zigzag phosphorene nanoribbons in three different widths of 12, 16, and 20 atoms. Based on the work in Ref. [27], in smaller ZPNR widths, the degeneracy of metallic localized edge states can be lifted due to interactions between the edge states.

The direct way to study the electronic properties of nanosystems and even large-scale structures is the tight-binding method. The tight-binding parameters can be obtained using a number of approaches. One of the most efficient and recent methods is the use of machine-learning interatomic potentials (MLIPs). The MLIPs can also be used to study the piezoelectric and mechanical properties of nanomaterials [28,29].

We use the five-parameter TB model of Ref. [30], fitted the energy bands near the Fermi energy. The model was introduced for the bulk two-dimensional black phosphorous. However, it has been experimentally verified in systems with non-perfect lattices, in particular in crystals with vacancies [31]. To obtain the transport coefficient (transmission probability), the non-equilibrium Green's function is used. Our numerical results indicate that the way the leads are connected to the system has a pronounced impact on the electron transmission probability. Also, because of the difference between the leads and the device edges, the metallic states of the leads, located on the edges of the ZPNRs, cannot be transmitted through the device, which makes the transport system behave as a semiconductor. The transport energy gap is mainly determined by the bulk bands of the nanoribbons of the leads. In addition, wider leads can capture smaller changes in the device section.

The manuscript is organized as follows. In the next section, the model and the method are presented. Results and discussion come in the third section, and finally, the results are summarized in the last section.

**Model and Method:**

In this work, we aim to investigate the electronic transport properties of phosphorene rings in two perpendicular directions (zigzag and armchair directions), in the presence of metallic zigzag nanoribbon leads, used as a source and drain contacts.

We have considered the model in such a way that the zigzag electrodes are connected to the two main edge geometries of the device, i.e., the armchair and zigzag sides of the ring, see Figure 1. A schematic of the models studied is presented in Figure 1. In panel (a), the electron transport goes along the zigzag direction of the lattice, labeled as configuration Z ($C_Z$). In panel (b), the transport direction is along the armchair direction, and we label it as configuration A ($C_A$).

The width of the leads is identified by the number of atoms in the width ($W_N$). Three different leads widths, $W_N$=12, 14, and 16 (N-ZPNR) were investigated in this work.

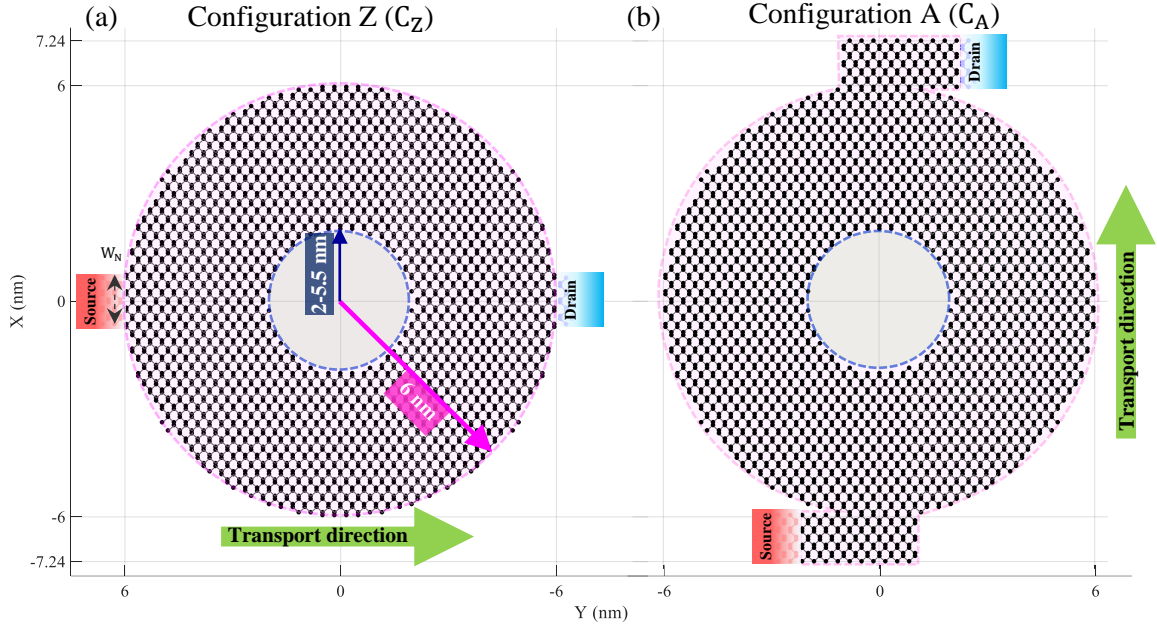

Figure 1. The studied model devices. (a) The two zigzag leads are connected to the device along each other in $C_Z$, (b) the case for two zigzag leads that are connected to the sides of the nanoring is labeled by $C_A$ (b).

The Hamiltonians are formed in the orthogonal one-orbital TB model with five parameters. It is shown that this model successfully reproduces the electronic properties of the phosphorene close to the Fermi energy [16,30,32]. We consider a system subject to the external potential that can be e.g. introduced by a tip of the scanning probe microscope.

The TB Hamiltonian reads:

$$H = \sum_i (\varepsilon_i |i\rangle\langle i| + V_i |i\rangle\langle i|) + \sum_{i,j}(t_{i,j}|i\rangle\langle j|+H.c.), \qquad (1)$$

where $\varepsilon_i$ is the on-site energy, $V_i$ is the external potential on the $i^{th}$ atom, and $t_{i,j}$ is the interatomic (hopping) parameter. The external potential is considered by $V_i = V_0 \left( \sqrt{\frac{x^2}{a} + \frac{y^2}{b}} - R_o \right)$ with a and b as control parameters related to the shape of the external potential. For $\sqrt{\frac{x^2}{a} + \frac{y^2}{b}} > R_o$, the external potential term, $V$, becomes zero. The profile of the external potential can be manipulated by the tip geometry, see [33]. We consider $V_0 = 0.1$ eV, and $R_o$=5.8 nm. The ratio of $\alpha = \frac{a}{b}$, is the parameter that determines the shape of the external potential.

The TB parameters are adopted from [16,32], with $\varepsilon_i = 0$ eV, and $t_1 = -1.22, t_2 = 3.665, t_3 = -0.205, t_4 = -0.105$, and $t_5 = -0.055$ eV for the essential five interactions. To obtain the electronic dispersion, one can solve the eigenvalue problem, as described in [18].

The TB Hamiltonians are then implemented in the NEGF formalism in order to study the electronic transport properties. The retarded Green's function can be evaluated as [34,35]:

$$G(E) = [(E + i\eta)\mathbf{I} - H_C - \Sigma_{SC}(E) - \Sigma_{DC}(E)]^{-1}, \qquad (2)$$

where $E$ is the electron energy, $\mathbf{I}$ is the identity matrix, $\eta$ is an arbitrarily small positive number, $H_C$ is the device Hamiltonian, and $\Sigma_{SC(DC)}$ is the self-energy for the source (drain) lead. Details about this formalism can be found in [35,36].

The spectral density operator is given by:

$$\Gamma_{S(D)}(E) = i[\Sigma_{SC(DC)}(E) - \Sigma_{SC(DC)}(E)^\dagger], \qquad (3)$$

The electron transmission probability can be obtained as follows:

$$T_e(E) = \text{Trace}[\Gamma_S(E)G(E)\Gamma_D(E)G(E)^\dagger]. \qquad (4)$$

The transport energy gap is evaluated in each of the structures. The energy gap is calculated by finding the first nonzero value with respect to zero energy in the transmission spectrum (assuming $E_F = 0$ eV). Additionally, the energy gap of the device is calculated by solving the eigenvalue problem of the device Hamiltonian, the first two significant energy differences are calculated, and the first states close to these differences are classified as energy domains that determine the energy gap of the isolated device [38].

Besides, the local density of states (LDOS) for a given atom (indicated by index *j*) is the imaginary part of the Green's function [37]:

$$\text{LDOS}(E)_j = \frac{-1}{\pi}\Im(G(E)_{j,j}), \qquad (5)$$

**Results and Discussion:**

The electronic transmission coefficient and transport energy gap for various inner radii (from 2 to 5.5 nm with a step of 0.1 nm) of a phosphorene ring with an outer radius of 6 nm are investigated using the TB model. The outer radius was fixed as the one beyond which its value has not a pronounced effect on the transport gap. The effect of the outer radius is investigated for a nanodisk without the central opening (or $R_{inner}=0$) with the zigzag configuration and the lead width of 12 atoms (Figure 2), which shows that the transport energy gap is insensitive to the larger radius. The results for other lead widths are presented in the Supplementary Information. The phosphorene is an intrinsic semiconductor, and the metallic behavior of the ZPNRs originates from the zigzag edge states. In circular geometries hosting various chiralities and in the limit of larger outer radii (where the transport energy gap converges to a value), the role of the outer edge in the electronic transport near the Fermi energy, or in energy ranges which are not captured by the electronic bands of the leads, is negligible.

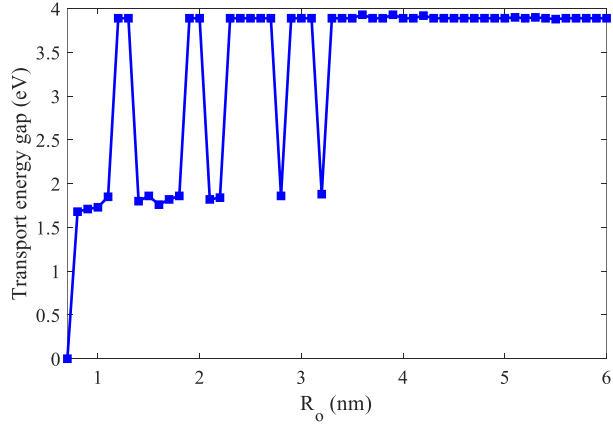

Figure 2. The effect of the outer radius (the case of $R_{inner}=0$) on the transport energy gap for $C_ZW_{12}$.

In the next step, in Figure 3, we plot the band structure and transmission spectrum as a function of energy, for three different lead widths together with the transmission spectrum for two different lead connections with an inner radius of 2 nm. Figure 3 shows the results for the zigzag phosphorene nanoribbons with $W_{12}$, $W_{16}$ and $W_{20}$ in (a), (b), and (c), respectively. As can be seen in Figure 3, with increasing width, the first two bulk bands (magenta bands) get closer to the Fermi energy. Therefore, the band gap converges to the intrinsic gap of the phosphorene for wider ribbons.

The transmission spectrum of both systems in all lead widths, i.e., $W_{12}$, $W_{16}$, and $W_{20}$ with $R_{inner}$ =2 nm, suggests that the transport energy gap is determined by the bulk bands of the ZPNRs, or on the other hand, the transport energy gap can be tuned by the width of the ZPNR up to the limits of the leads. In the smallest ZPNR, the first two bulk bands (marked by magenta color) are far from the Fermi energy. However, as the width increases, they get closer to each other until they reach the intrinsic energy gap of the phosphorene. The transport gap and its size are essential in determining some physical properties, such as the Seebeck coefficient [18,39].

We plot the whole transmission spectrum in the energy, but one should note that the five-parameter TB model is fitted so that energy bands close to the Fermi energy are justified (gray transparent shaded areas gives a schematic in Figure 3). As a general rule, we notice that the transmission coefficient is closer to the maximum value set by the zigzag leads in configuration Z than in configuration C for all widths, which is a sign of the intrinsic anisotropy of phosphorene.

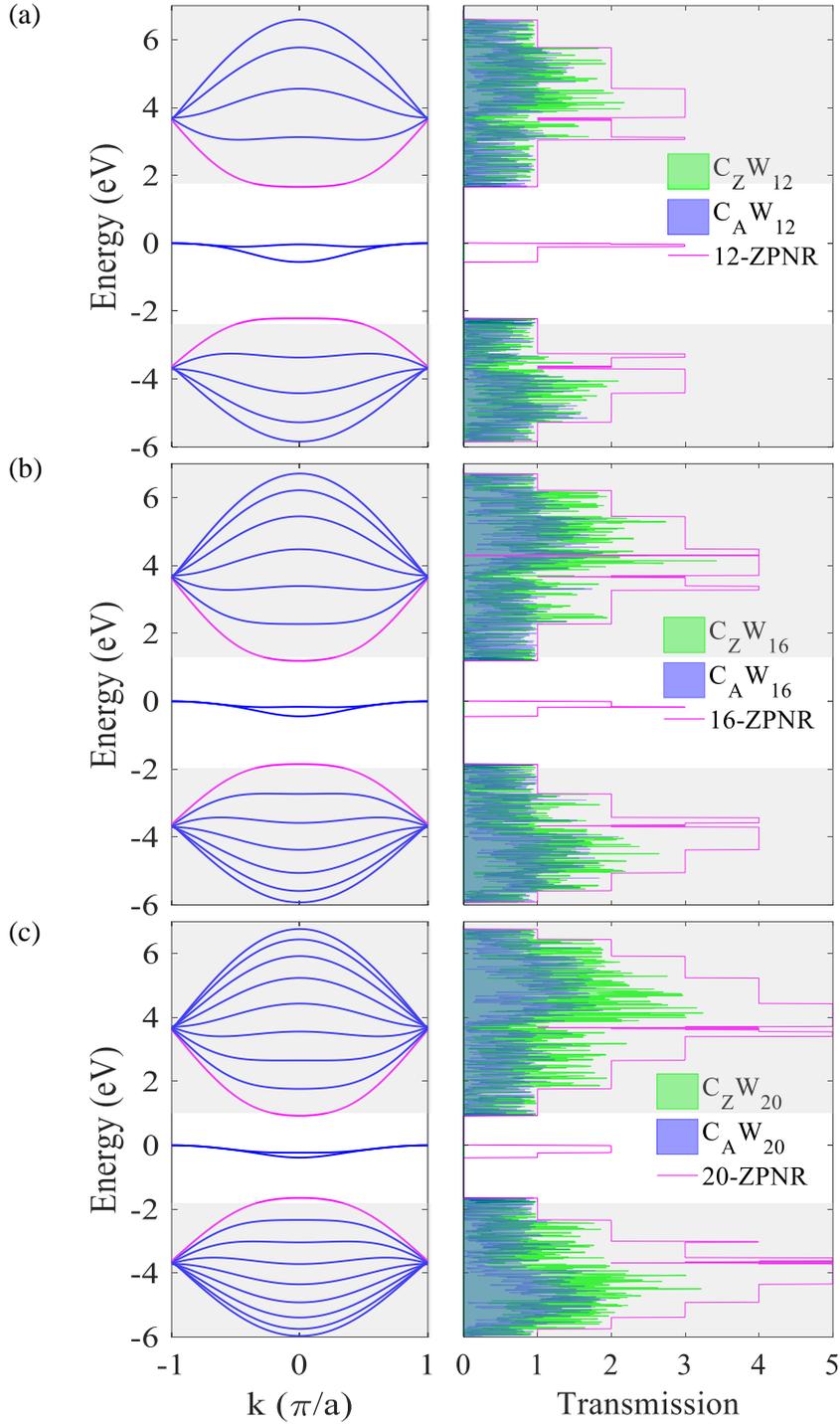

Figure 3. The electronic band structure and transmission spectrum of the systems, (a) 12-ZPNR, (b) 16-ZPNR, and (c) 20-ZPNR. Secondary electronic bands (here bulk bands of the ZPNR) with respect to the Fermi energy are shown by magenta lines in the band structure plots. The transmission spectrum of the ribbons is shown in magenta, the zigzag configuration is shown in green, and the armchair configuration is illustrated in blue-magenta.

In the next step, we studied the transport energy gap for the systems as a function of the inner radius. We find that the energy stays almost constant up to a certain inner radius (Figure 4 (a)), and then it changes drastically. As noted earlier, by increasing the width of the ZPNR, the bulk bands get close to the Fermi energy; the leads provide a broader energy range that can then be filtered by the finite conductance of the device. This explains why the difference between transport energy gaps for different lead widths is significant. As the lead width increases, the range of the transport gap modulation by the inner radius also increases, indicating that wider leads are advantageous for sensing applications. Therefore, in addition to the width of the leads, the inner radius of the nanoring can be used to tune the transport energy gap.

Figure 4 (b) shows that a dependence of the energy gap of the $C_A$ configuration is close to that of the $C_Z$ system. Since the outer radius is constant, this close behavior indicates the dominant role of the inner edge. However, a slight difference can still be noticed between the panels of Figure 4 (a) and (b), which can be attributed to the response of the inner edge to different transport directions.

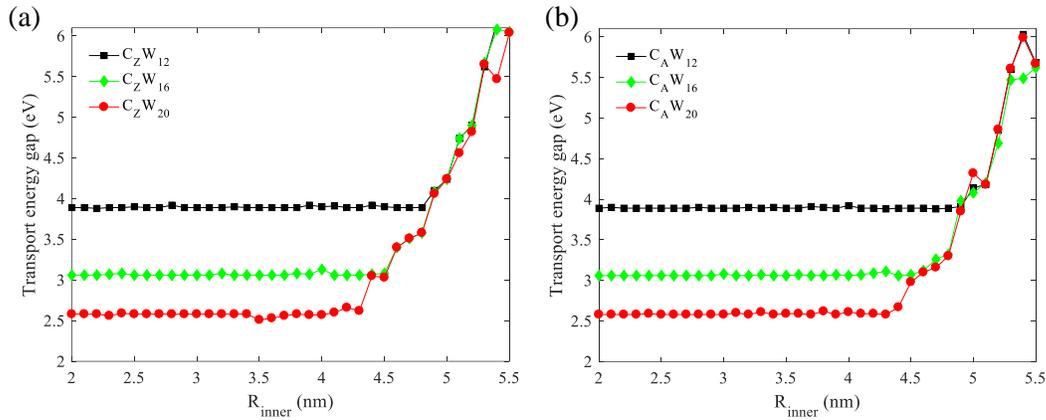

Figure 4. Transport energy gap as a function of inner radius for various lead widths for $C_Z$ (a), and $C_A$ (b).

The energy gap of the isolated ring is plotted in Figure 5, showing an increasing trend in the value of the gap as the inner radius increases. As discussed in Ref. [40], charges in the armchair phosphorene nanoribbons are more localized in the central part of the ribbon, while in ZPNRs, they are localized on the edge. As one can see, by increasing the inner radius or equivalently, reducing ring width, the number of zigzag edge atoms increases, which in turn affects the energy gap.

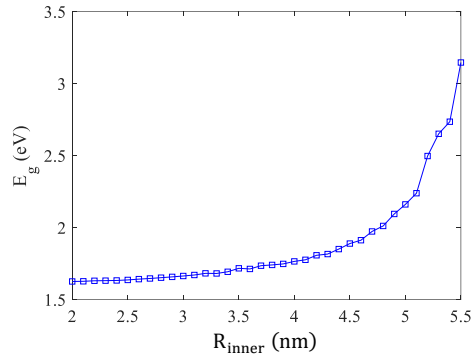

Figure 5. Isolated device (quantum ring) energy gap as a function of the inner radius. The outer radius is 6 nm.

The energy levels, together with the square of the wavefunction (probability amplitude) of an isolated phosphorene nanoring, without the connection to the leads, are shown in Figure 6 (a) and (b) for two radii of 2 and 5 nm, respectively. We emphasize that the energy gap of the quantum ring is different from the one which it is connected to leads, forming a transport system. The probability amplitude is mapped onto the geometry of the quantum dot. We present a probability densities for the energy range of the isolated device, including $|\Psi_1|^2$ which is related to the state at the edge of the energy gap above the Fermi energy, $|\Psi_2|^2$, which is related to a typical state within the energy gap , and $|\Psi_3|^2$ which belongs to the state at the edge of the energy gap below the Fermi energy. The energy eigenvalues of confined states are the closest energies to the Fermi energy.

The in-gap states are almost localized at the edges, similar as the low-energy edge states of a zigzag phosphorene nanoribbon. Comparing the cases of the inner radius of 2 and 5 nm shows that for the thinner ring ($R_{inner}$=5 nm) confined states get closer to each other and their distribution in the width of the ring becomes smoother along the edges. For a periodic system, like ZPNRs, these states can form nearly flat bands, see Figure 3. As the width decreases, their bands become more dispersive.

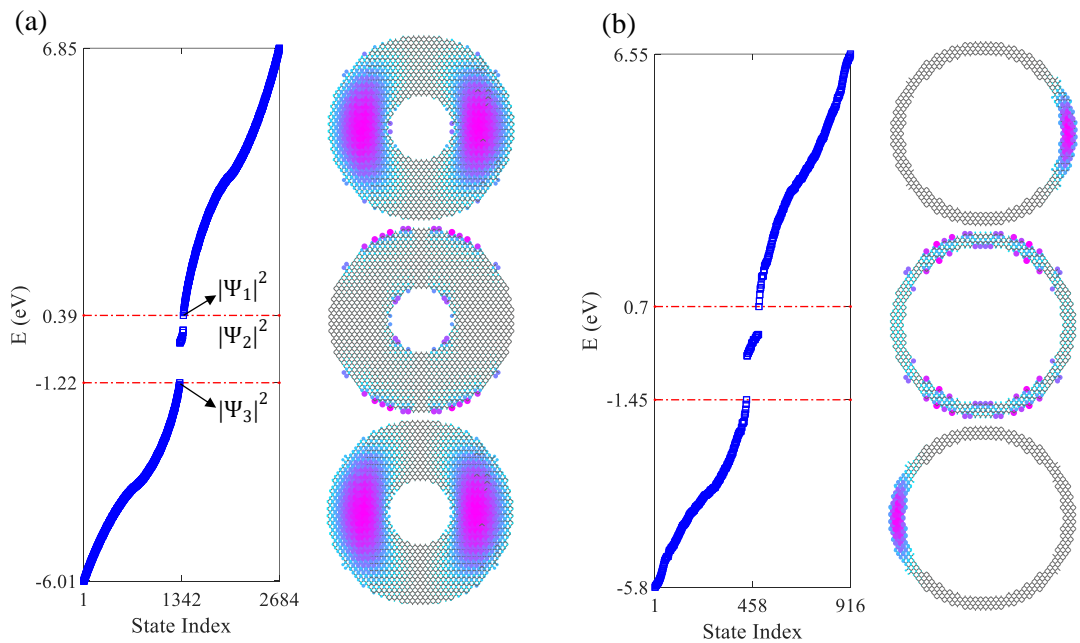

Figure 6. (a) and (b) are energy levels of the device section for an inner radius of 2 and 5 nm together with a representation of $|\Psi|^2$ for three types of the states (see the text).

The estimate of the isolated device energy gap (molecular energy gap) cannot be directly translated to the one of the devices used for transport, with the leads attached. In the NEGF method, leads are assumed to be infinite, i.e., the device becomes equilibrated with leads through the interface of the device and leads. The coupling of the quantum ring and electrodes can change the energy levels of the system. The ZPNR leads are metallic. However, the metallic edge states cannot propagate through the device, as the edge is modified within the quantum ring. In order to locate the areas within the ring that carry the current fed by ZPNR, one can study the local density of states. The LDOS shows how many states are available for electrons in a particular energy. The results are presented in the Supplementary Information.

Let us now study the tunability of the transport properties of the device with the external potential. The effect of an external potential (with $V_0 = 0.1$ eV) is studied in Figure 7. The effect of an external potential with various $\alpha$ is more pronounced in $C_Z$ than in $C_A$. For an external potential with an ellipse shape that stretched along the zigzag direction ($\alpha$=0.5, green line), the transport coefficient of the conduction band is close to unity. In the valence band, the transmission value is more suppressed compared to the case with $V=0$ (magenta line). Moreover, the transport gap of the $C_Z$ is more sensitive to this type of external potential than $C_A$, especially for $\alpha=1$, one can see that the transport gap extends by about 0.35 eV (for $C_A$ this change is about 0.04 eV).

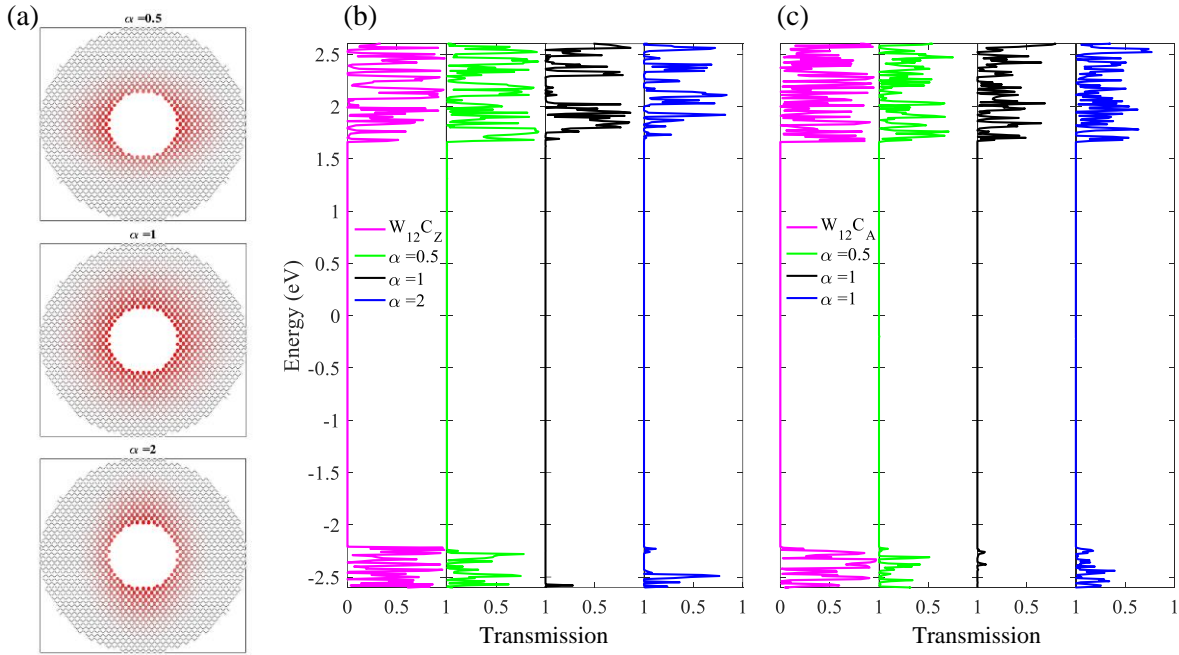

Figure 7. Map of three external potentials added as on-site energy to a nanoring, (a) with an inner radius of 2 nm. The zero potential is marked by gray color and the maximum potential by the most intense red color. The transmission spectra for (b) $W_{12}C_Z$, (c) $W_{12}C_A$, for three external potentials shown in panel (a).

## Conclusion

We have studied the electronic transport properties of phosphorene nanorings with a fixed outer radius of 6 nm, and a range of inner radii of 2-5.5 nm. Two configurations were considered for attaching metallic ZPNR leads with widths of 12, 16, and 20 atoms to study the effect of lead width on the transport properties. Electronic transport properties were studied with the help of the five-parameter TB model implemented in the NEGF formalism. Based on the numerical results, the effect of the outer radius disappears at large radii, e.g., in the case of $C_Z W_{12}$, for $R_o \geq 3.3$ nm, the transport energy gap remains almost constant.

The results show that all of the structures were semiconductors whose transport energy gaps are determined by (i) the inner radius of the ring, and (ii) either the conduction or valence band side of the Fermi energy, i.e., the bands associated with the bulk of the nanoribbon, not the edge. The transport energy gap shows more sensitivity for the case of wider leads due to their richer electronic configuration. This indicates that wider leads may be useful for sensing applications. Also, the intrinsic anisotropy of the transport within the phosphorene is translated into a difference in the transmission spectrum of two configurations of the leads. The transmission coefficient suppression for $C_A$ is larger than $C_Z$. The circular shape of the external potential (with $V_0 = 0.1$ eV), widens the transport gap about 0.35 eV, and 0.04 eV in $C_Z$ and $C_A$, respectively.

Farhad Khoeini, khoeini@znu.ac.ir

Supplementary Information:

# Insights into electronic and transport properties of phosphorene nanorings in two perpendicular directions: Effects of circular and elliptical external potentials

M.Amir Bazrafshan[1], Farhad Khoeini*,[1], Bartlomiej Szafran[2]

[1]*Department of Physics, University of Zanjan, P.O. Box 45195-313, Zanjan, Iran*

[2]*Faculty of Physics and Applied Computer Science, AGH University of Science and Technology, al. Mickiewicza 30, 30-059 Kraków, Poland*

*Corresponding author: Farhad Khoeini, khoeini@znu.ac.ir

The effect of the outer radius is investigated for a nanodisk without the central opening (or $R_{inner} = 0$) with the zigzag configuration and the lead width of 16 and 20 atoms in Figure.S. 1 (a) and (b), respectively. The radius of the disks is not smaller than the width of the ribbon, i.e., the smallest disk radius makes a perfect lead.

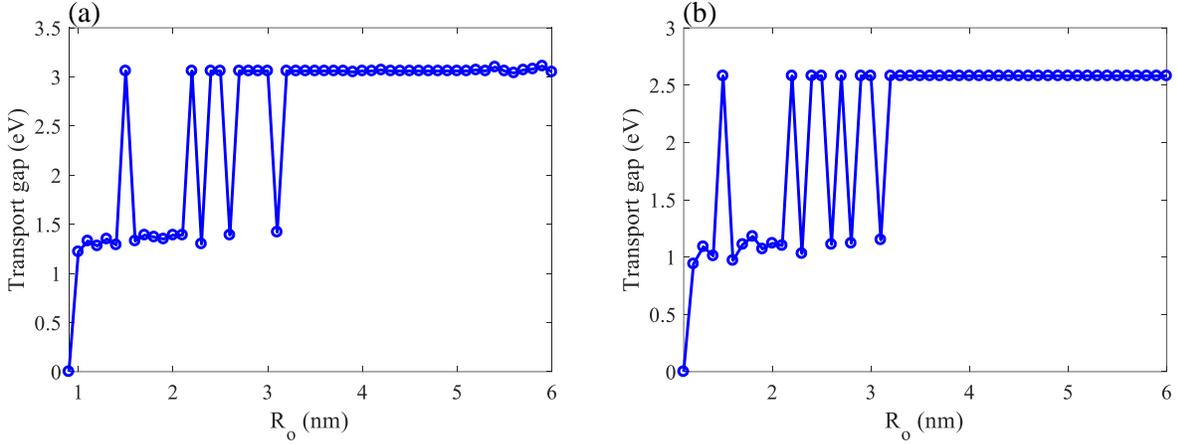

Figure.S. 1 The effect of the outer radius (the case of $R_{inner}=0$) on the transport energy gap for $C_ZW_{16}$ (a), and $C_ZW_{20}$ (b).

The LDOS of the isolated ring at the Fermi energy (E=0 eV) with $R_{inner}$=2 nm shown in Figure S.2 (a) exhibits a single radial line with maximal values along the zigzag crystal direction and two perpendicular maxima oriented along the armchair direction.

Upon connection of the electrodes in the $C_ZW_{12}$ configuration the LDOS [Figure S.2 (b)] becomes more uniform with a trace of the maxima near the inner edge of the ring along the lines oriented in the armchair direction. For the leads attached for the armchair transport direction $C_AW_{12}$ [Figure S.2 (c)], we observe a single maximal line oriented radially in the armchair crystal direction, with a larger visibility than for the $C_Z$ configuration (see the scale of the color bar). The LDOS of the rings connected in the $C_Z$ and $C_A$ configurations are plotted in panels (b) and (c) of Figure S.2, respectively. Attachment of the leads in $C_Z$ [Figure S.2 (b)] and [Figure S.2 (c)] reduces the ocalization of the LDOS with respect to the case of the isolated ring (see the scale of the color bar).

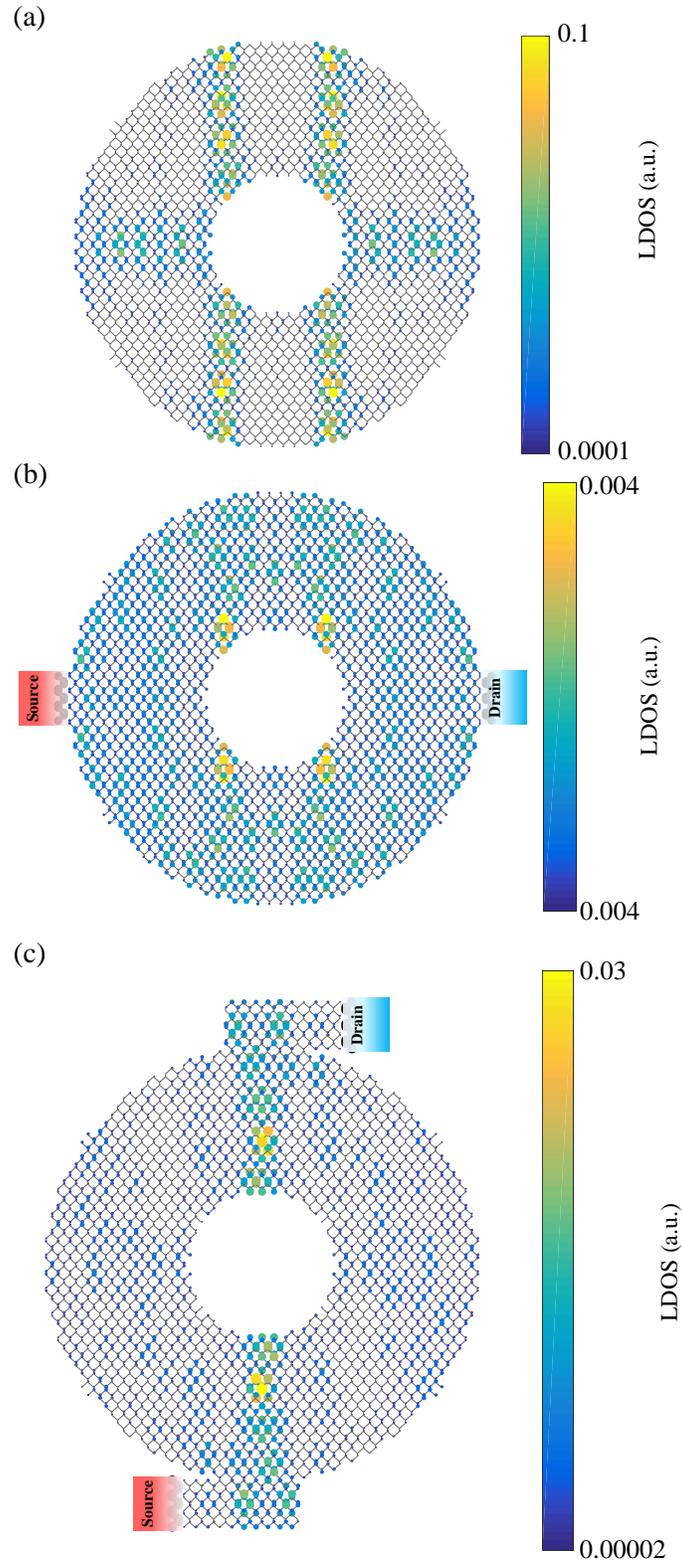

Figure S.8. Local density of states at E=0 eV for the phosphorene system with (a)$R_{inner}$ = 2 nm quantum ring, (b) $C_ZW_{12}$, and (c) $C_AW_{12}$.